\documentclass[12pt]{article}
\usepackage{graphicx}
\begin{document}

\centerline{\bf Nonequilibrium Zaklan model on Apollonian networks$^{\star}$ }

\bigskip

\centerline{ F. W. S. Lima}

\bigskip

\noindent

{\it Dietrich Stauffer Computational Physics Lab}, Departamento de F\'{\i}sica,\\

Universidade Federal do Piau\'{\i}, 64049-550, Teresina - PI, Brazil.\\

\medskip

e-mail: fwslima@gmail.com

\bigskip

\hspace{4.7pc}{$\star$ This paper is dedicated to Dietrich Stauffer}\\

\begin{abstract}

The Zaklan model had been proposed and studied recently using the equilibrium
Ising model on Square Lattices (SL) by Zaklan et al (2008), near the critica
temperature of the Ising model presenting a well-defined phase transition; but
on normal and modified Apollonian networks (ANs), Andrade et al. (2005, 2009)
studied the equilibrium Ising model. They showed the equilibrium Ising model not
to present on ANs a phase transition of the type for the 2D Ising model. Here,
using agent-based Monte-Carlo simulations, we study the Zaklan
model with the well-known majority-vote model (MVM) with noise and apply it to
tax evasion on ANs, to show that differently from the Ising model the MVM on ANs
presents a well defined phase transition. To control the tax evasion in the
economics model proposed by Zaklan et al, MVM is applied in the neighborhood of
the critical noise $q_{c}$ to the Zaklan model. Here we show that the Zaklan
model is robust because this can be studied besides using equilibrium dynamics
of Ising model also through the nonequilibrium MVM and on various topologies
giving the same behavior regardless of dynamic or topology used here.
\end{abstract}

\bigskip

\bigskip

\section{Introduction}

The Ising model \cite{a3,onsager} has become a excellent tool to study other
models of social application. Therefore, following this line of reasoning the
Zaklan model had been proposed and studied recently using the equilibrium Ising
model on square Lattices by Zaklan et al. \cite{zaklan,zaklan1,lima8}. Lima
\cite{limanew}, based on Grinstein et al. \cite{g}, made a proposal to extend
the current model (Zaklan's model) to nonequilibrium systems, using
nonequilibrium Majority-Vote Model (MVM) \cite{mario} in order to make Zaklan's
model more realistic, because tax evasion is nonequilibrium.

Our simulation is based on the well-known Apollonian packing introducing
Apollonian networks \cite{soares}.
According to Andrade et al. \cite{soares}
the ANs are simultaneously scale-free \cite{barabasi}, small-world \cite{dukan},
Euclidean, space filling, and with matching graphs \cite{boyd}. Therefore, the
ANs have social connections that are often similar to scale-free or small-world
networks \cite{fortunato} and have been studied e.g. for the Ising model and a 
magnetic
model \cite{andrade1,andrade2}. The ANs here are defined and described in detail
in the pioneering work by Andrade et al. \cite{soares}.

As shown in \cite{soares,andrade1,andrade2, Nuno} on ANs, the
Ising and Potts models do not present a phase transition. Therefore, the
Ising model on this topology is not useful for the Zaklan model, because it does
not have a phase transitions on ANs.
Therefore, our work is not only to show that the Zaklan model works another a
topology as the ANs, but also to
show that in some topologies the traditional equilibrium spin models, as
the Ising model, are not appropriate to study socio-economic models
as the model proposed by Zaklan et al \cite{zaklan,zaklan1}. Therefore,
different from Ising models, the nonequilibrium MVM model is presented here
as an alternative model in the study of socio-economic Zaklan model on ANs. This
similar behavior was also shown by Sumour et al \cite{sumuor} and Lima and
Stauffer \cite{lima-stauffer}, where the Ising model do not present a phase
transition on directed Bab\'asi-Albert networks and hypercubic lattices,
respectively. Thus, we present an
alternative proposed in the study of social and economics models via
nonequilibrium MVM and also showed that in this case the criterion of Grinstein
et al. \cite{g} is not applicable, i.e, on ANs Ising and MVM do not belong to 
the same universality class.

Our paper is organised as follows. In section 2, we present the
Zaklan model evolving with dynamics of MVM. In section 3 we make an analysis of
tax evasion dynamics with the Zaklan model on ANs, using
MVM for their temporal evolution under different enforcement regimes; we discuss
the results obtained. In section 4 we show that MVM also is capable to control
the different levels of the tax evasion analysed in section 3, as it was made by
Zaklan et al. \cite{zaklan1} using Ising models. We use the enforcement
mechanism cited above on ANs and discuss the resulting tax evasion dynamics.
Finally in section 5 we present our conclusions about the study of the Zaklan
model using MVM on ANs.

\section{Zaklan model and evolution dynamics}

\subsection{\bf Zaklan model}

The Zaklan model \cite{zaklan} consists of agents located on a
regular or irregular structure.  Each agent is represented by an individual 
spin $S_{i} = \pm 1$, who can either be an honest tax payer $+1$ or a cheater 
$-1$. Initially everybody is assumed honest. In each iteration individuals can 
rethink their behaviour and have the opportunity to become the opposite type 
of agent they were in the previous period. Each agent's environment may prefer 
tax evasion or reject it. The agent depends on two factors: 
First, the agent's environment exerts influence on the agent in the next period.
Second, people's decisions are partly autonomous, independent of their 
environment, This autonomous part is responsible for the emergence
of tax evasion, because some initially honest tax payers decide to
evade taxes and then exert influence on others to do so as well. 
Tax evaders have the greatest influence to turn honest citizens into tax 
evaders if they constitute a majority in the respective neighbourhood. 
On the other hand, if most people in the vicinity are honest, the respective 
individual is likely to become a tax payer if (s)he was a tax evader before.

The model also presents an enforcement mechanism that consists of two
components: a probability of an efficient audit $p$. If tax evasion is detected,
the individual must remain honest for a number $k$ of periods to be specified.
One iteration is one sweep through the entire lattice. The temporal evolution
this model can be performed by using an equilibrium or, in the present work, by
nonequilibrium dynamics.

\subsection{\bf Zaklan model via nonequilibrium MVM model}

The Apollonian networks contain $N=3+(3^{n-1}-1)/2$ nodes (sites, agents, spins)
 where $n$ is the generation number \cite{andrade1}.  Our MVM dynamics 
contains a noise parameter $q$. For the case of four neighbors and 
$q=0$, if three or four neighbors disagree with the central site, the center 
flips; if one or none disagrees, the center does not  flip; if two agree
and two disagree, it flips with probability 1/2. If $q > 0$ the center 
may disobey this majority rule. More precisely,
in each time period the system evolves by a single spin-flip
dynamics with a probability $w$ given by

\begin{equation}
w(\sigma_i)=\frac{1}{2}\biggl[ 1-(1-2q)\sigma_{i}S\biggl(\sum_{\delta
=1}^{k_{i}}\sigma_{i,\delta}\biggl)\biggl],
\end{equation}

where $S(x)$ is the sign $\pm 1$ of $x$ if $x\neq0$, $S(x)=0$ if $x=0$, and the
summation runs over all $k_i$ nearest-neighbour sites $\sigma_{i,\delta}$ of
$\sigma_{i}$. In this model an agent assumes the value $\pm 1$ depending on the
opinion of the majority of its neighbors. The noise parameter $q$ plays
the role of the temperature in equilibrium systems and measures the probability
of aligning $\sigma_{i}$ antiparallel to the majority of its neighbors
$\sigma_{i,\delta}$.

\section{Controlling the tax evasion dynamics}


\begin{figure}[ht]
\begin{center}
\includegraphics[angle=0,scale=0.4]{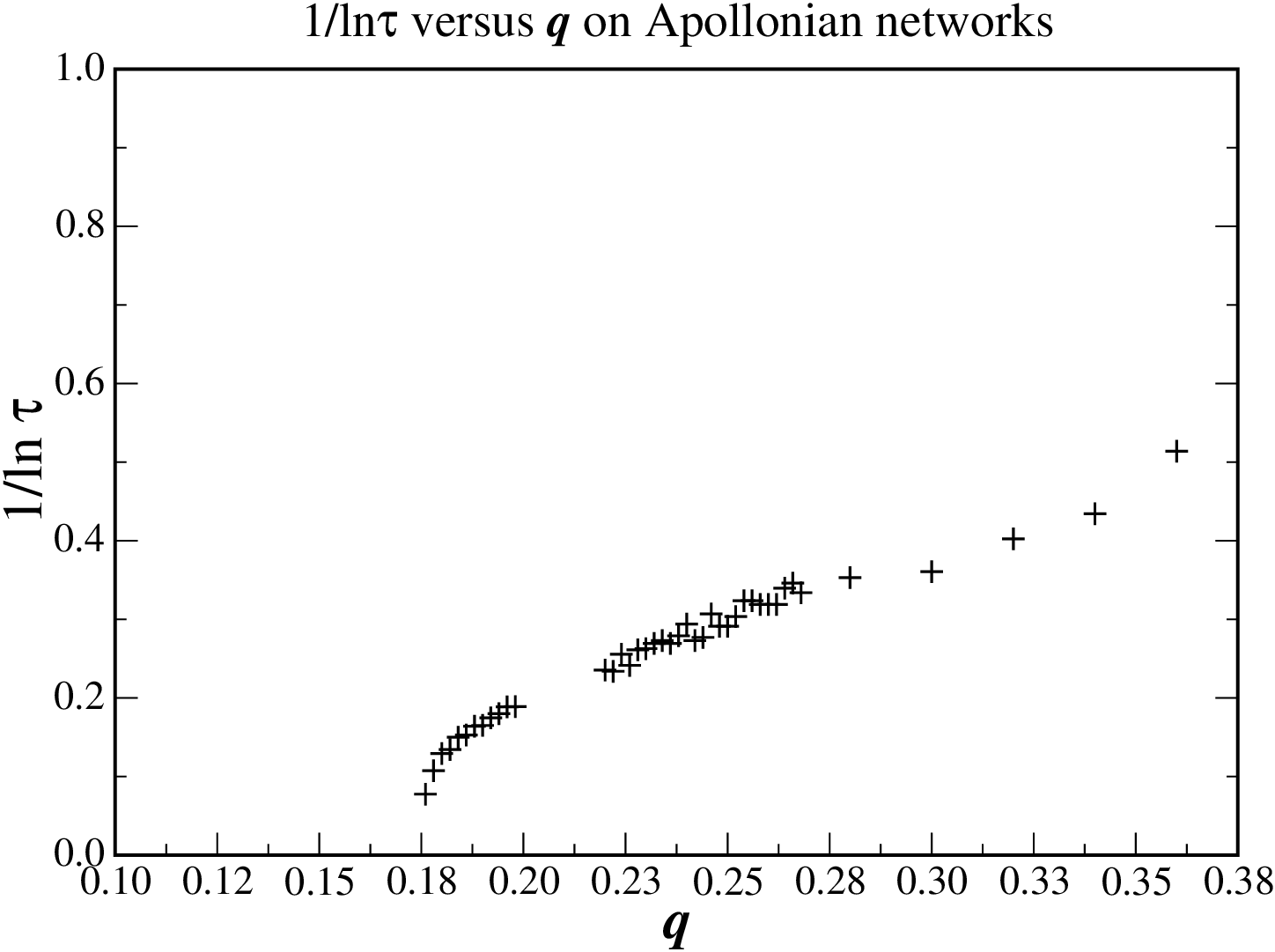}
\end{center}
\caption{Reciprocal logarithm of the relaxation times on ANs for versus $q$.}
\end{figure}

In order to test if there is a phase transition in MVM models on ANs, we
measured the relaxation time $\tau$ as a funtion of the noise
parameter $q$, independent of our tax question. We start the system
with all spins up and a number $N$ of spins equal to $7,174,456$ ($n=16$). We
determine
the time $\tau$ after which the magnetisation $\sum_i \sigma_i$ has flipped
its sign for the first time, and then take the median value of nine
samples. As one can see in Fig. 1, the relaxation time goes to
infinity at some positive $q$ value near 0.18, indicating a second order phase
transtion. On contrast, the Ising model on ANs \cite{andrade1,andrade2} and
directed BA networks has no
phase transition and agrees with the modified Arrhenius law for
relaxation time \cite{sumuor}. In order to improve earlier affirmation the
magnetisation and
susceptibility for $N = 3,283; \; 9,844; \; 29, 527; \; 88,576$, and 
$265,723$ sites and with $n = 8, 9,$ $10, 11,$ and $12$ generation are plotted 
in Fig. 2 \cite{apollo}.

\begin{figure}[ht]
\begin{center}
\includegraphics[angle=0,scale=0.25]{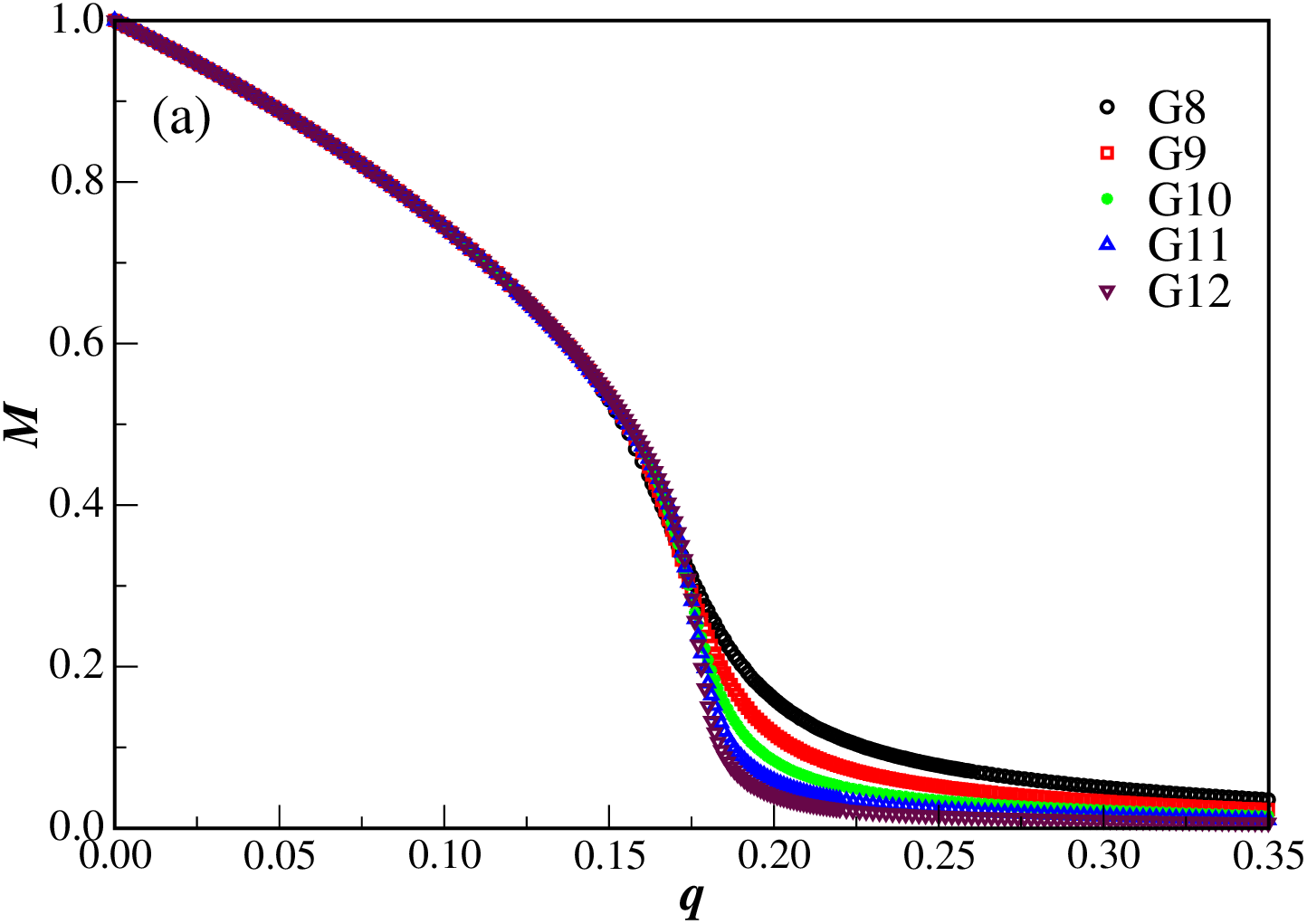}
\includegraphics[angle=0,scale=0.25]{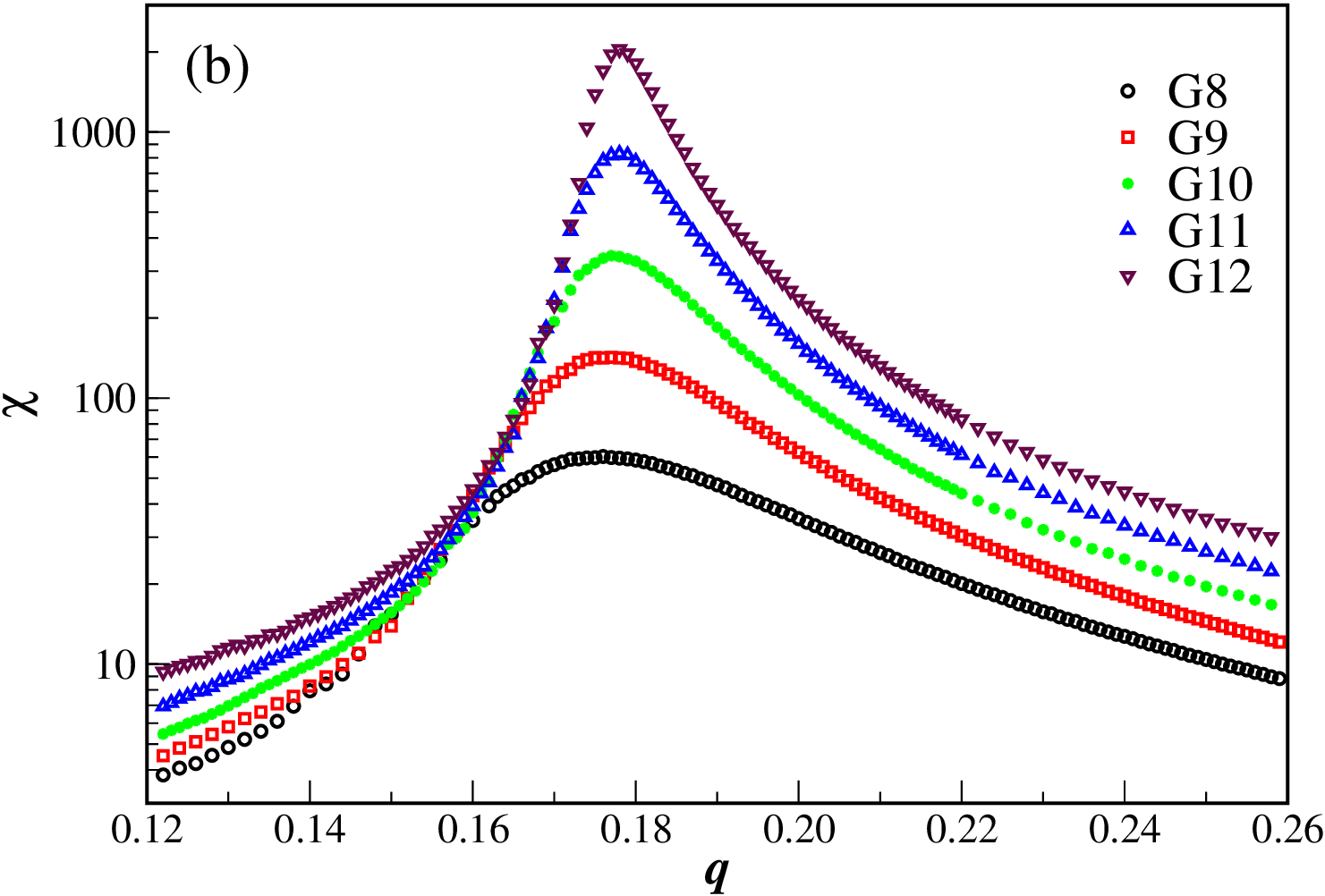}
\end{center}
\caption{Plot of the magnetisation (a) and susceptibility (b) for different
generations of the Apollonian Network,(circle) $G8$, (square) $G9$, (star)
$G10$, (triangle up) $G11$, (triangle down) $G12$. The number after $G$ gives
the generation number $n$.}
\end{figure}
\bigskip

In order to calculate the rate of tax evaders, we use

\begin{equation}
{\rm tax \; evasion} =\frac{[N-N_{\rm honest}]}{N},
\end{equation}
where $N$ is the total number and $N_{\rm honest}$ the honest number of agents.
The tax evasion is calculated at every time step $t$ of system evolution; one
time step is one sweep through the entire network.

Here, we follow the same steps we did in a previous work \cite{limanew}.
Therefore, we first will present the
baseline case $k=0$ and $p=0.0\%$, i.e., no use of enforcement, at $q=0.80q_{c}$
and with $N=367 \; (n=7)$ sites
for ANs. All simulation are performed over $25,000$ time steps, as shown in Fig.
3. For very low noise the part of autonomous decisions almost completely
disappears. The individuals then base their decision solely on what most of
their neighbours do. A rising noise has the opposite effect. Individuals then
decide more autonomously. Therefore, Figure 3 was expanded to four examples,
in order to show how much the results differ if one changes the random numbers.
Error bars cannot describe this randomness properly. (For the later figures the
error bars are visible from the fluctuations in time which show a band of
fractions.) Although everybody is honest initially, it is impossible to predict
roughly which level of tax compliance will be reached at some time step in the
future.

For MVM it is known that for $q > q_{c}$, half of the people are honest and the
other half cheat, while for $q<q_{c}$ either one opinion or the other opinion
dominates. Because of
this behavior we set at fixed "Social Temperature" ($q$) to some values slightly
below $q_{c}$, where the case that agents distribute in equal proportions onto
the two alternatives is excluded. Then having set the noise parameter $q$ close
to $q_{c} \simeq 0.18$ on the ANs, we vary the degrees of punishment ($k=1$,
$10$ and $50$) and audit probability
rate ($p=0.5\%$, $10\%$ and $90\%$). Therefore, if tax evasion is detected by 
the enforcement mechanism $p$, the period of punishment $k$ is triggered in
order to control the tax evasion level. The punished individuals
remain honest for a certain number $k$ of periods, as explained before in
section 2.
\bigskip

\begin{figure}[ht]
\begin{center}
\includegraphics[angle=0,scale=0.45]{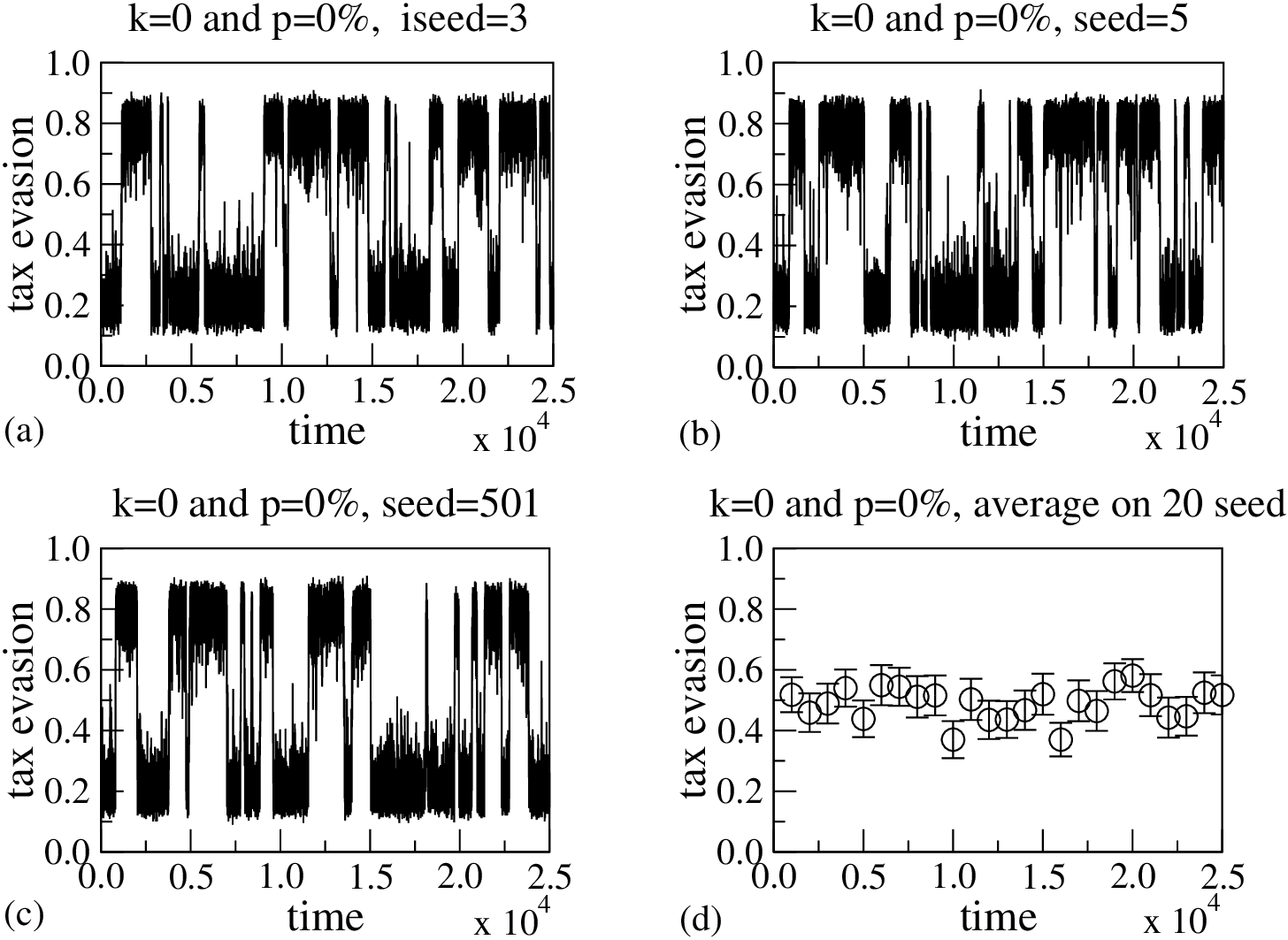}
\end{center}
\caption{Baseline case:$k=p=0$ and with three different seed (a), (b), (c) and
the average over twenty different seeds (d). We use $q=0.80q_{c}$ on ANs and
perform all simulations over $25,000$ time steps, also in the later figures.}
\end{figure}

Figure 4 illustrates different simulation settings for ANs, for
each considered combination of degree of punishment ($k=1$, $10$ and $50$)
and audit probability ($p=0.5\%$, $10\%$ and $90\%$), where the tax evasion
is plotted over $25,000$ time steps. Both a rise in
audit probability (greater $p$) and a higher penalty (greater $k$) work to
flatten the time series of tax evasion and to shift the band of possible
non-compliance values towards more compliance. However, the simulations
show that even extreme enforcement measures ($p=90\%$ and $k=50$)
cannot fully solve the problem of tax evasion.

\begin{figure}[hbt]
\begin{center}
\includegraphics[angle=0,scale=0.5]{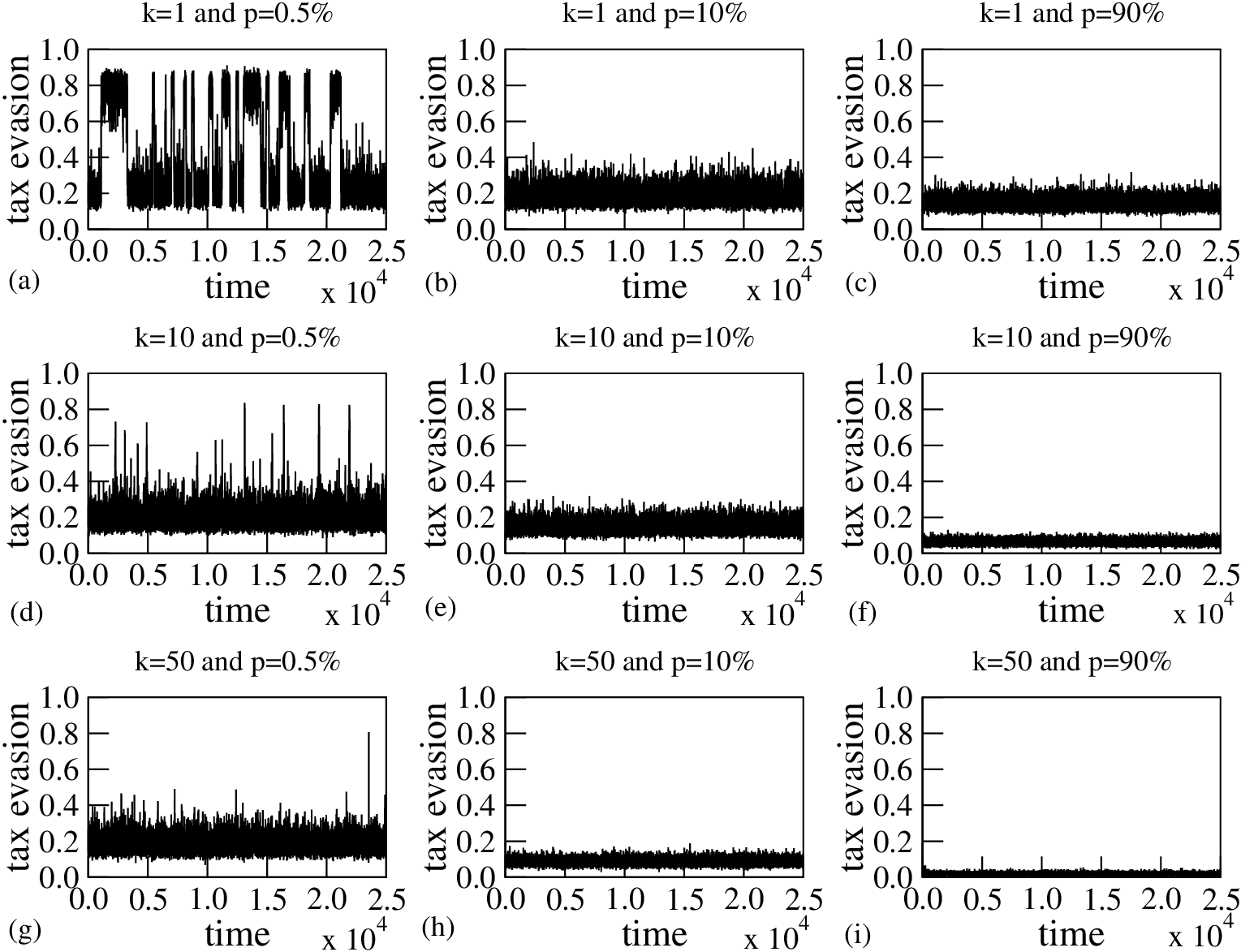}
\end{center}
\caption{Tax evasion for different enforcement regimes ANs and for degrees of
punishment $k=0$, $1$, $10$ and $50$ and audit
probability $p=0.5\%$, $10\%$ and $90\%$. }
\end{figure}

In Fig. 5 we plot tax evasion for ANs, but now with $N=3,283$, again for
different enforcement $k$ and audit probability $p$. Now the fluctuations are
much smaller since the network is nearly nine times larger. For case (a) we plot
the baseline case $k=0$ and $p=0$, i.e., no use of enforcement for ANs and
parameters as in Fig. 3. The probable error for part (c) fluctuates near 0.0031
and is much smaller than the symbols (circle). Case (b) with $k=1, \; p=0.5\%$
shows already a strong reduction of tax evasion. In case (c) we show the tax
evasion level decreases, on ANs, for a more realistic set of possible values
degrees of punishment $k=10$ and audit probability $p=4.5\%$ \cite{WGa, zaklan}.
In case (d) we also show the tax evasion level decreases much more for an
extreme set of punishment $k=50$ and audit probability $p=90\%$ \cite{zaklan}.
Therefore, our model also works for large networks.
\newpage

\begin{figure}[ht]
\begin{center}
\includegraphics[angle=0,scale=0.5]{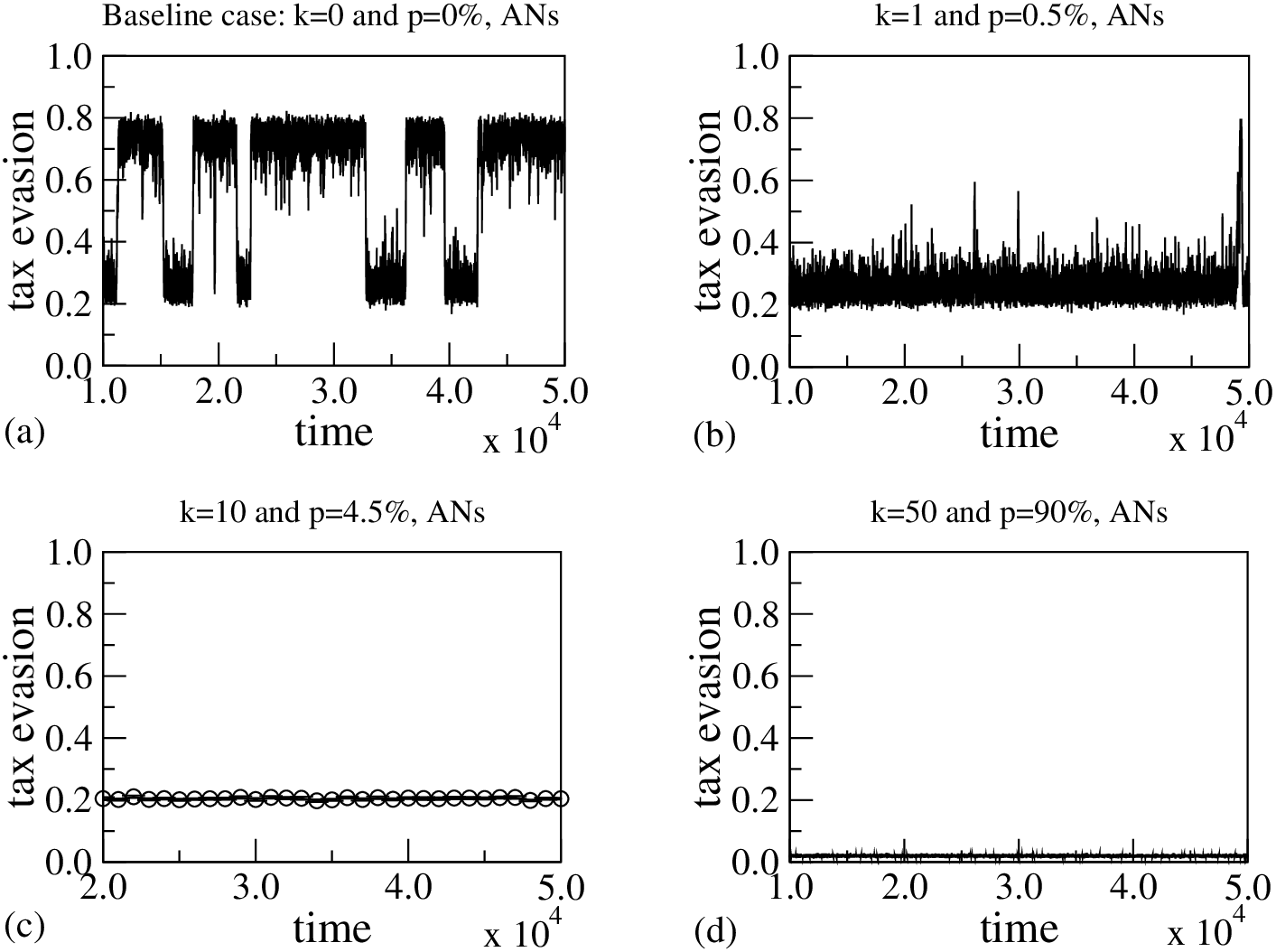}
\end{center}
\caption{ Tax evasion for different enforcement regimes ANs and for degrees of
punishment $k=0$, $1$, $10$ and $50$ and audit probability
$p=0.0\%, 0.5\%, 4.5\%$, and $90\%$ for $N=3,283$
sites (nodes) of ANs and use $50,000$ time steps. Here, for $k=0$ and $p=4.5\%$
(c), we present the average over twenty different seeds. }
\end{figure}

To understand statistical errors, in Fig. 5 we plot tax evasion for ANs with
$N=3,283$ now for the case $k=10$ and $p=4.5\%$.  We found from 20 samples in
part (c) that the tax evasion remains at around
$20\%$, but with fluctuations in time larger than from sample to sample: The
probable errors are much smaller than the fluctuations seen in part (c).
\bigskip
\bigskip
\newpage
\section{Conclusion}

Less developed countries may have more tax evasion because of less trust in 
government \cite{WGa}. To study this problem Zaklan et al. 
\cite{zaklan,zaklan1} proposed
a model, called here the Zaklan model, using Monte Carlo simulations and a
equilibrium dynamics (Ising model) on square lattices. Their results are in good
agreement with analytical and experimental results obtained by \cite{WGa}. In
this work we show that the Zaklan model of tax evasiom is very robust
because we use a nonequilibrium dynamics (MVM) to simulate the
Zaklan model, with results similar to equilibrium dynamics
(Ising model) \cite{zaklan,zaklan1}, and also on various topologies
\cite{limanew}. Also here we found the plausible result that tax evasion is 
diminished by higher audit probability $p$ and stronger punishment $k$. 

\bigskip

\bigskip

{\bf Acknowledgments}

The author thanks D. Stauffer for many suggestion and fruitful discussions
during the development this work and also for the reading this paper. We also
acknowledge the Brazilian agency CNPQ for its financial support. This
work also was supported the system SGI Altix 1350 the computational park
CENAPAD.UNICAMP-USP, SP-BRAZIL.

\end{document}